\documentclass[useAMS,usenatbib]{mn2e}
\usepackage{epsfig,amsmath,natbib}
\def\be{\begin{equation}} 
\def\ee{\end{equation}}

\def\mpc{\,{\rm {Mpc}}}

\def\gsim{\lower.5ex\hbox{\gtsima}} 
\def\lsim{\lower.5ex\hbox{\ltsima}} \def\gtsima{$\; \buildrel > \over 
\sim \;$} \def\ltsima{$\; \buildrel < \over \sim \;$} \def\prosima{$\; 
\buildrel \propto \over \sim \;$} \def\gsim{\lower.5ex\hbox{\gtsima}} 
\def\lsim{\lower.5ex\hbox{\ltsima}} 
\def\simgt{\lower.5ex\hbox{\gtsima}} 
\def\simlt{\lower.5ex\hbox{\ltsima}} 
\def\simpr{\lower.5ex\hbox{\prosima}} \def\la{\lsim} \def\ga{\gsim} 
  
\def\gtsima{$\; \buildrel > \over \sim \;$} 
\def\ltsima{$\; \buildrel < \over \sim \;$} 
\def\gsim{\lower.5ex\hbox{\gtsima}} 
\def\lsim{\lower.5ex\hbox{\ltsima}} 
\def\simgt{\lower.5ex\hbox{\gtsima}} 
\def\simlt{\lower.5ex\hbox{\ltsima}} 
\def\simpr{\lower.5ex\hbox{\prosima}} 
\def\la{\lsim} 
\def\ga{\gsim} 
\def\Zcr{Z_{\rm cr}}

\def\Msun{M_{\odot}} %{\,{\rm M_\odot}}
\def\Zsun{Z_{\odot}} 
\def\Lsun{L_{\odot}}

\def\E3{{\cal E}_{\rm g}^{III}}

\title{Metals and ionizing photons from dwarf galaxies}
\author[Salvadori, Tolstoy, Ferrara, Zaroubi]
{S. Salvadori$^{1}$, E. Tolstoy$^{1}$, A. Ferrara$^{2}$, S. Zaroubi$^{1}$\\
$^1$Kapteyn Astronomical Institute, Landleven 12, 9747 AD Groningen, The
Netherlands\\
$^2$Scuola Normale Superiore, Piazza dei Cavalieri 7, 56126 Pisa, Italy}
\begin{document} 
\date{} 
\pagerange{\pageref{firstpage}--\pageref{lastpage}} \pubyear{} 
\maketitle 
\label{firstpage} 
\begin{abstract}
%%%%%%%%%%%%%%%%%%%%%%%%%%%%%%%%%%%%%%%%%%%%%%%%%%%%%%%%%%%%%%%%%%%%%%%%%%%
We estimate the potential contribution of $M<10^9\Msun$ dwarf galaxies to 
the reionization and early metal-enrichment of the Milky Way environment, 
or circum-Galactic Medium. Our approach is to use the observed properties 
of ancient stars ($\gsim12$ Gyr old) measured in nearby dwarf galaxies to 
characterize the star-formation at high-$z$. We use a merger-tree model 
for the build-up of the Milky Way, which self-consistently accounts for 
feedback processes, and which is calibrated to match the present-day 
properties of the Galaxy and its dwarf satellites. We show that the 
high-$z$ analogues of nearby dwarf galaxies can produce the bulk of 
ionizing radiation ($>80\%$) required to reionize the Milky Way environment. 
Our fiducial model shows that the gaseous environment can be $50\%$ 
reionized at $z\approx8$ by galaxies with $10^7\Msun\leq{M<10^8\Msun}$. 
At later times, radiative feedback stops the star-formation in these 
small systems, and reionization is completed by more massive dwarf 
galaxies by $z_{rei}=6.4\pm0.5$. The metals ejected by supernova-driven
outflows from $M<10^9\Msun$ dwarf galaxies almost uniformly fill the 
Milky Way environment by $z\approx5$, enriching it to $Z\approx 2\times 
10^{-2}\Zsun$. At $z\approx2$ these early metals are still found to
represent the $\approx50\%$ of the total mass of heavy elements in
the circum-Galactic Medium.
%%%%%%%%%%%%%%%%%%%%%%%%%%%%%%%%%%%%%%%%%%%%%%%%%%%%%%%%%%%%%%%%%%%%%%%%%%%%%%%%
\end{abstract}
%%%%%%%%%%%%%%%%%%%%%%%%%%%%%%%%%%%%%%%%%%%%%%%%%%%%%%%%%%%%%%%%%%%%%%%%%%%%%%%
\section{Introduction}
%%%%%%%%%%%%%%%%%%%%%%%%%%%%%%%%%%%%%%%%%%%%%%%%%%%%%%%%%%%%%%%%%%%%%%%%%%%%%%%%
The nature of the first astrophysical sources that reionized 
the Inter-Galactic-Medium (IGM) by $z\leq6$ \citep{fan2006}, and 
possibly controlled its metal-enrichment up to $Z\geq10^{-4}\Zsun$ 
\citep{songaila2001,ryanweber2009} is still a matter of debate. Recent models 
and observations of cosmic reionization require an increasing UV 
emissivity towards high-redshifts, and suggest that undetected 
faint galaxies, $M_{UV}>-18$, must have significantly contributed 
to reionization \citep[e.g.][]{bolton2007,salvaterra2011,
alvarez2012,bouwens2012,finkelstein2012,kuhlen2012,mitra2013}.
Supernova-driven outflows from dwarf galaxies are likely to be 
one of the most efficient mechanisms to transport metals into 
the IGM \citep[e.g.,][]{maclow1999,madau2001,scannapieco2002}, 
making these systems the obvious candidates for polluting it. 
However, the physical properties of high-$z$ dwarfs are very 
uncertain, and their formation is predicted to be progressively 
suppressed during reionization by the increased UV radiation field 
intensity \citep[e.g.][]{ciardiferrara2005}.

Ancient ($\gsim12$~Gyr) metal-poor stars observed in nearby dwarf 
spheroidal galaxies (dSphs) offer the unique opportunity to look 
back at the star-formation properties of these small systems during 
the epoch of reionization \citep[][hereafter SF09]{salva2009}. 
These ancient stars were formed before the gravitational interaction 
with the Galaxy could have substantially altered the evolution of 
dSphs \citep[e.g.,][]{ibata2001,penarrubia2008}. Thus their observed 
features, reflect the intrinsic evolution of high-$z$ dwarfs, 
along with the (possible) influence of dissociating/ionizing radiation. 
Nearby dSphs have been studied in great detail, and an ancient stellar 
population has been observed in all of them \citep[e.g.,][]{tolstoy2009}. 
The Sculptor dSph is a typical example of "classical" dSphs, which have
total luminosity $L\approx(10^5-10^{7.5})\Lsun$ and total mass $M\approx
(10^8-10^9)\Msun$, and stars older than $12$~Gyr represent $\gsim55\%$ 
of its total stellar mass \citep{deboer2012a}. Ultra-faint 
dwarf galaxies, $L < 10^5\Lsun$ and $M<10^8\Msun$, seem to show no evidence 
for stars younger than $12.5$~Gyr \citep{okamoto2012,dallora2012,brown2012}, 
suggesting that the star-formation activity might have been suppressed 
by a global event, such as reionization.
 
In this {\it Letter} we focus on the high-redshift ($z>5$) 
analogues of Milky Way (MW) dwarf galaxies to
investigate the impact of these systems on the reionization 
{\it and} metal-enrichment of the MW environment, the formation 
medium of the Galaxy and its dwarf satellites. We use the 
data-calibrated cosmological code GAMETE (GAlaxy MErger Tree 
and Evolution, SF09, \cite{salva2007} and \cite{salva2012}, 
hereafter SSF07 and SF12), which has been developed to investigate 
the properties of present-day ancient metal-poor stars, and which 
successfully reproduces the metallicity-luminosity 
relation of MW dwarf galaxies (SF09), the stellar Metallicity 
Distribution Function (MDF) observed in the Galactic halo (SSF07), 
in classical, and in ultra-faint dwarf galaxies (SF09). 
\begin{figure}   
\begin{centering}
  \includegraphics[width=0.95\columnwidth]{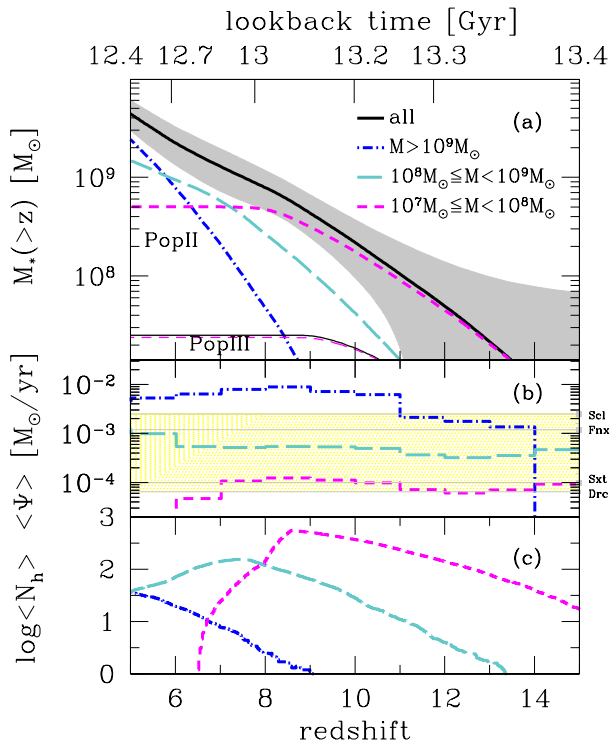}
    \caption{(a) The cumulative stellar mass, $M_*(>z)$, (solid line), 
    (b) the mean SF rate, $\langle\Psi\rangle$, and (c) the mean number 
    of star-forming haloes, $\langle N_h\rangle$, as a function of $z$. 
    The lines show the contribution to these physical quantities by haloes 
    within different mass ranges given in (a). The results are averaged over 
    50 MW merger histories, and the ${\pm}1\sigma$ dispersion is indicated 
    by the shaded area in (a). The Pop~II/Pop~III contribution 
    to $M_*(>z)$ is shown in (a). 
    The shaded area in (b) shows the range of $\langle\Psi\rangle$ measured 
    in MW classical dSphs, the horizontal lines are measurements for: 
    Sculptor \citep{deboer2012a}, Fornax \citep{deboer2012b}, 
    Sextans \citep{lee2009}, Draco \citep{aparicio2001}.}
    \label{Fig1}
\end{centering}
\end{figure}
%%%%%%%%%%%%%%%%%%%%%%%%%%%%%%%%%%%%%%%%%%%%%%%%%%%%%%%%%%%%%%%%%%%%%%%%%%%%%
\section{Data-calibrated model}    
%%%%%%%%%%%%%%%%%%%%%%%%%%%%%%%%%%%%%%%%%%%%%%%%%%%%%%%%%%%%%%%%%%
GAMETE reconstructs possible merger histories of a MW-sized dark matter 
halo, $M_{mw}=10^{12}\Msun$, by using a Monte Carlo algorithm based 
on the Extended Press-Schechter theory (SSF07). The star formation 
(SF) is traced along the merger trees by adopting physically 
motivated hypotheses: (a) there exists a minimum halo mass to form 
stars, $M_{sf}(z)$, whose evolution accounts for the suppression of 
SF in progressively {\it more massive} objects due to the increasing 
photo-dissociating/ionizing radiation (Fig.~1 of SF09),
(b) the SF rate is proportional to the mass of cold gas in each galaxy, 
and it is regulated via the SF efficiency, $\epsilon_*$, 
(c) in {\it minihaloes} with virial temperatures $T_{vir}\simlt10^4$~K, 
the SF efficiency is reduced as 
$\epsilon_{H_2}=2{\epsilon_*}[1+\big(T_{vir}/{2\times10^4K}\big)^{-3}]^{-1}$ 
to account for the ineffective cooling by molecular hydrogen, H$_2$ (SF09), 
(d) Population~II (Pop~II) stars form according to a Larson Initial Mass 
Function (IMF, SSF07) if the metallicity of the gas exceeds the critical 
value, $\Zcr\approx10^{-5\pm1}\Zsun$, which sets the minimal conditions
to trigger the formation of low-mass stars \citep{schneider2002}; 
if $Z<Z_{cr}$ then Population~III (Pop~III) stars form, with the IMF 
discussed in Sec.~3.

The chemical evolution of the gas is simultaneously traced in the haloes 
and in the surrounding MW environment by including the effect of SN-driven
outflows, which are controlled by the SN-wind efficiency, $\epsilon_w$ 
(SF12). 
The total volume filled by metals, $V_{tot}(z)$, is computed at each $z$ 
as the sum of the volumes of individual metal bubbles around star-forming 
haloes, $V_s^{i}(z)$. The time evolution of each bubble is obtained by 
solving the canonical momentum and energy conservation equations derived 
adopting the thin-shell approximation (Madau et al. 2001, eqs. 24-25). 
This can be translated into a volume filling factor, $Q_Z$, given by:
\be
Q_Z=\frac{V_{tot}(z)}{V_{mw}(z)}=\frac{4\pi}{3V_{mw}(z)}\Sigma_{i}\big[R_s^i(z)\big]^3
\ee
where $R_s^i(z)$ is the shell radius, whose expansion is driven by the 
energy input of the ensamble of SNe exploding in each star-forming halo.
The proper MW volume, $V_{mw}(z)$, is estimated at the turn-around radius, 
$V_{mw}(z)\approx M_{mw}/[5.55\Omega_m\rho_{cr}(z)]\approx5~\textrm{Mpc}^3(1+z)^{-3}$.

The SF and SN wind efficiencies are fixed to reproduce 
the global properties of the MW (SSF07). The low-$Z$ tail of the Galactic 
halo MDF allows us to constrain $Z_{cr}\leq 10^{-4}\Zsun$, here fixed to 
the best fitting value, $10^{-4}\Zsun$ (SSF07). The reconstructed evolution 
of $M_{sf}(z)$, correctly matches the observed metallicity-luminosity 
relation of MW dwarf galaxies (SF09), and it implies a reionization history 
that is consistent with the early/late reionization models of 
\cite{gallerani2006} (see Fig.~1 of SF12). 
%%%%%%%%%%%%%%%%%%%%%%%%%%%%%%%%%%%%%%%%%%%%%%%%%%%%%%%%%%%%%%%%%%%%%%%%%
%\begin{figure}   
%\begin{centering}
%  \includegraphics[width=0.91\columnwidth]{FIG1}
%    \caption{(a) The cumulative stellar mass, $M_*(>z)$, (solid line), 
%    (b) the mean SF rate, $\langle\Psi\rangle$, and (c) the mean number 
%    of star-forming haloes, $\langle N_h\rangle$, as a function of $z$. 
%    The lines show the contribution to these physical quantities by haloes 
%    within different mass ranges given in (a). The results are averaged over 
%    50 MW merger histories, and the ${\pm}1\sigma$ dispersion is indicated 
%    by the shaded area in (a). The Pop~II/Pop~III contribution 
%    to $M_*(>z)$ is shown in (a). 
%    The shaded area in (b) shows the range of $\langle\Psi\rangle$ measured 
%    in MW classical dSphs, the horizontal lines are measurements for: 
%    Sculptor \citep{deboer2012a}, Fornax \citep{deboer2012b}, 
%    Sextans \citep{lee2009}, Draco \citep{aparicio2001}.}
%    \label{Fig1}
%\end{centering}
%\end{figure}
%%%%%%%%%%%%%%%%%%%%%%%%%%%%%%%%%%%%%%%%%%%%%%%%%%%%%%%%%%%%%%%%%%%%%%%%%%%%%
\subsection{Ionizing photon production}
%%%%%%%%%%%%%%%%%%%%%%%%%%%%%%%%%%%%%%%%%%%%%%%%%%%%%%%%%%%%%%%%%%%%%%%%%%%%%
The cumulative rate of ionizing photons produced by stellar populations 
with different ages and metallicities is computed along the merger trees 
by including the results of population synthesis models in GAMETE. We 
use {\tt STARBURST99}\footnote{http://www.stsci.edu/science/starburst99} 
\citep{leitherer1999} to obtain the evolution of the ionizing photon 
production rate per stellar mass formed, $q(t,Z)$, during each burst
of star-formation with $Z=[0.02-1]\Zsun$. Because of the limited 
metallicity range we use $q(t,Z_{min})$ for bursts of Pop~II stars 
with $Z\leq0.02\Zsun$, and we follow the results of \cite{schaerer2003} 
to compute $q(t,0)$ for Pop~III stars. At each $z$ the rate of ionizing 
photons per unit volume injected into the MW environment, ${\dot n}_{\gamma}$, 
is then computed as:
\be
{\dot n}_{\gamma}(z)=V_{mw}^{-1}\Sigma_{i}\,[f_{\rm esc}\,q(t_z-t_{z_i},Z)\,M_*(z_i,Z)]\;, 
\label{eq:N_gamma}
\ee
where $M_*(z_i,Z)$ is the mass of stars with metallicity $Z$ formed at $z_i$, 
and $f_{\rm esc}$ is the escape fraction of ionizing photons, which may   
depend on the nature of the stellar sources and $z$. The evolution of the 
filling factor of ionized regions, $Q_{HII}$, is obtained by integrating:
\begin{equation} 
\frac{dQ_{HII}}{dt}=\frac{{\dot n_{\gamma}}}{n^{mw}_H}-\frac{Q_{HII}}{t_{rec}}
\label{eq:QHII}
\end{equation}
where $n^{mw}_H$ is the comoving hydrogen number density of the MW
environment, and $t_{rec}=[{C\alpha_B\,n^{mw}_H\,(1+z)^3}]^{-1}$ 
is the hydrogen recombination time, with $C$ clumping factor, and 
$\alpha_B=2.6\times10^{-13}$cm$^3$s$^{-1}$ recombination rate. 
We estimate $n^{mw}_H=M_{gm}(z)/(\mu m_p V_{mw})$, as the ratio 
between mass of gas in the MW environment, $M_{gm}(z)\approx(\Omega_b
/\Omega_m)[M_{mw}-M_{coll}(z)]$, and the comoving MW volume, where 
$M_{coll}(z)$ is the total mass in collapsed haloes.
We find $n^{mw}_H\approx(5.4-4.2)\,n^0_H$ for $z=(20-5)$, where 
$n^0_H$ is the hydrogen number density in the IGM.
%%%%%%%%%%%%%%%%%%%%%%%%%%%%%%%%%%%%%%%%%%%%%%%%%%%%%%%%%%%%%%%%%%%%%%%%%%%%%%%
\begin{figure*}
  \includegraphics[width=1.55\columnwidth]{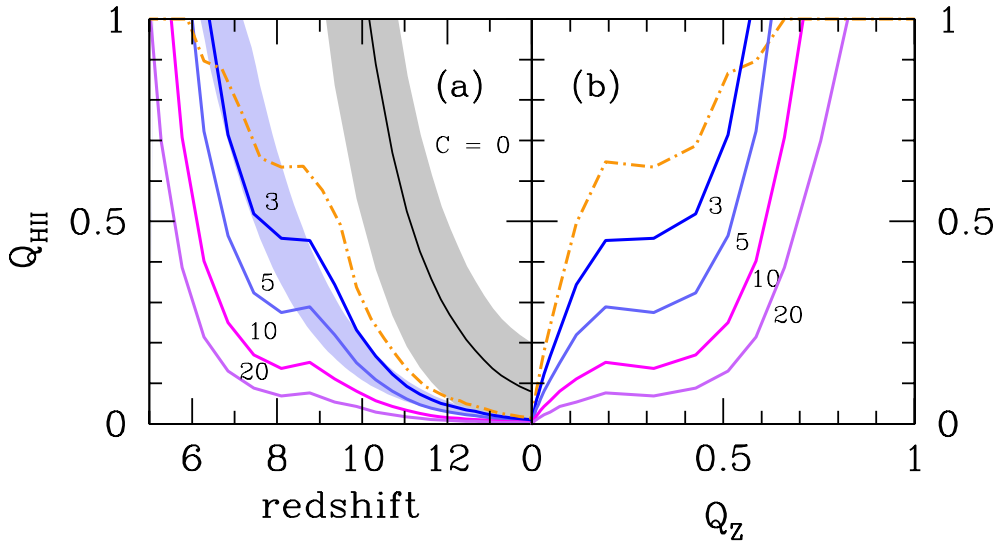}
   \caption{(a) the filling factor of ionized regions in the MW volume,
    $Q_{HII}$, at different $z$. The {\it solid curves} show the results
    for $f_{esc}=0.1$ and for increasing clumping factors, $C$, (from top
    to bottom). The {\it dot-dashed curve} shows the $Q_{HII}$ evolution
    obtained by assuming $C=4$ and $f_{esc}(z)$ \citep{mitra2013}. All
    results are averaged over 50 MW merger histories. The {\it gray shaded
    area} shows the $\pm 1 \sigma$ dispersion for the case $C=0$. The {\it
    blue shaded} area delimits the early/late reionization histories by
    \cite{gallerani2006}. (b) The volume filling factors of metals, $Q_Z$,
    vs ionized regions, $Q_{HII}$, for different $C$-$f_{esc}$ combinations.}
    \label{Fig2}
\end{figure*}
%%%%%%%%%%%%%%%%%%%%%%%%%%%%%%%%%%%%%%%%%%%%%%%%%%%%%%%%%%%%%%%%%%%%%%%%%%%%%
\section{Results} %{Present-day analogues of high-$z$ galaxies}
%%%%%%%%%%%%%%%%%%%%%%%%%%%%%%%%%%%%%%%%%%%%%%%%%%%%%%%%%%%%%%%%%%%%%%%%%%
The contribution to the total stellar mass, $M_*(>z)$, of haloes within 
different mass ranges is shown in Fig.~1a as a function of redshift. 
We can see that the evolution of $M_*(>z)$ is dominated by haloes 
with increasing $M$ for decreasing $z$, which is a consequence 
of hierarchical galaxy formation. Those haloes with $M<10^9\Msun$, 
represent the dominant contributors until $z\approx6$, when 
$M_*(>z)\approx2\times10^9\Msun$. Pop~III stars form in $M<10^8\Msun$ 
objects, and already at $z<15$ they are sub-dominant with respect 
to Pop~II stars. However, to avoid overestimating the photons/metals 
that can be contributed by low-mass dwarf galaxies, we adopt a 
conservative hypothesis, and assume Pop~III stars to have standard 
IMF and $m_{PopIII}=(1,100) \Msun$ (see SSF07). 

The stellar mass required to reionize the MW environment, 
$M_*\geq{M_{gm}}(1+N_{rec})/(f_{esc}N_{\gamma})$, can be estimated         
by recalling that $N_{\gamma}\approx4000$ ionizing photons are produced per
baryon in stars with metallicity $Z\geq 1/20 \Zsun$ and distributed with 
a Salpeter IMF. Assuming that only $10\%$ of these photons escape galaxies, 
taking $M_{gm}\approx\Omega_b/\Omega_mM_{mw}$, and ignoring recombinations, 
$N_{rec}=0$, we get $M_*\geq4.25\times10^8\Msun$. This implies (Fig.~1a) 
that $M<10^9\Msun$ haloes can be responsible for the reionization of the 
MW environment, which is expected to occur at $z_{rei}\lsim 9$. So, what 
are the present-day analogues of these high-$z$ objects?

Galaxies with total masses $10^7\Msun\leq{M}<10^8\Msun$ formed at
$z\leq10$ are mostly minihaloes, $T_{vir}\propto{M}^{2/3}(1+z)<10^4$~K.
The ones that are not incorporated into larger galaxies are expected to
be observed today as ultra-faint dwarf galaxies (SF09). Their tiny
stellar masses ($10^3\Msun\lsim{M}_*<10^5\Msun$) reflect both their low
{\it mean SF rate}, $\langle\Psi\rangle$, and their short SF activity (Fig.~1b).
The low $\langle\Psi\rangle$ is due to the ineffective cooling by H$_2$
molecules and gas-loss, driven by SNe. The limited duration is a
consequence of reionization, as shown in Fig.~1c that displays the 
{\it average number of star-forming haloes}, $\langle{N}_h\rangle$,
for different mass ranges. At $z>9$ haloes with $M<10^8\Msun$ are 
$\approx10$ times more abundant than more massive objects at, but 
their $\langle{N}_h\rangle$ rapidly declines for $z<8.5$ due to the 
increased Lyman-Werner (LW) background that photo-dissociates H$_2$ 
in minihaloes, turning them into sterile systems (SF12). Following 
reionization gas accretion stops for $T_{vir}<2\times10^4$~K objects,
quenching the SF in the lowest-mass galaxies. These findings are in 
agreement with the {\it Hubble Space Telescopes} results for ultra-faint 
dwarfs SF histories \citep{brown2012}.

Galaxies with $10^8\Msun\leq M<10^9\Msun$, mainly form at
$z<11$ by the merging of less massive progenitors. 
They are expected to be the high-$z$ analogues of classical 
dSphs (SF09). These galaxies are not affected by LW
background, and continue forming stars after the end of
reionization ($z<6$, Fig.~1b) although they slowly decrease
in number (Fig.~1c) as they merge into more massive systems.
At $z>5$ the predicted $\langle\Psi\rangle$ for these galaxies
is consistent with the mean value measured for the oldest stellar
population ($>12.5$~Gyr) of the classical MW dSphs, as seen
in Fig.~1b.

Galaxies with $M>10^9\Msun$ assembled at $z<9$ by the merging 
of smaller galaxies (Fig.~1c). They are tipically associated 
to the major components of the MW, which at $z\approx 6$ have 
masses $M\approx10^{9.5}-10^{11}\Msun$ \citep{salva2010b}.

Now we can use our model to quantify the impact of the 
$z>5$ analogues of present-day dwarf galaxies, i.e. 
$M<10^9\Msun$ haloes, on the evolution of the MW environment.
%%%%%%%%%%%%%%%%%%%%%%%%%%%%%%%%%%%%%%%%%%%%%%%%%%%%%%%%%%%%%%%%%%%%%%%%%%%%%%%%
\subsection{Reionization sources}
%%%%%%%%%%%%%%%%%%%%%%%%%%%%%%%%%%%%%%%%%%%%%%%%%%%%%%%%%%%%%%%%%%%%%%%%%%%%%
\begin{figure}
  \includegraphics[height=0.85\columnwidth]{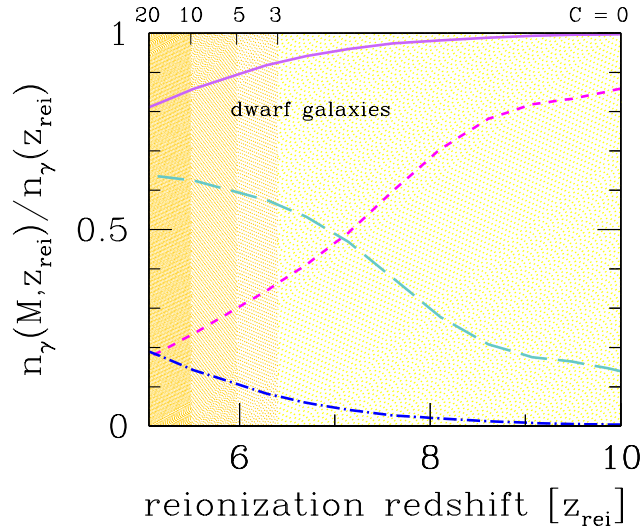}
  \caption{Time-integrated comoving photon density, $n_{\gamma}$,
    normalized to the total value at different reionization redshifts,
    $z_{rei}$. The mean contribution by haloes within different mass
    ranges is shown using the same lines as in Fig.~1a. The solid curve
    shows the cumulative contribution by $M<10^9\Msun$ dwarfs.
    The shaded regions associate $z_{rei}$ to the corresponding clumping
    factors, $C$.}
    \label{Fig3}
\end{figure}
The evolution of the volume filling factor of ionized regions, $Q_{HII}$, 
is shown in Fig.~2a for different values of the clumping factor $C$, and 
$f_{esc}=0.1$. The growth of $Q_{HII}$ is regulated by the balance between 
the rates of ionization, ${\dot N}_{\gamma}$, and recombination, 
${\dot N}_{rec}=n^{mw}_H\,V_{mw}\,t^{-1}_{rec}=\alpha_B\,C\,(n^{mw}_H)^2V_{mw}(1+z)^3$. 
If recombinations are ignored, i.e. $C=0$, then $Q_{HII}$ increases 
steeply, and the MW gets reionized at $z_{rei}\approx10$ ($Q_{HII}=0.99$). 
Reionization is delayed if larger $C$ values are assumed, although all 
models with $C>0$ are consistent with $z_{rei}\approx6$ within $2\sigma$ 
errors. At $z\geq10$ recombinations easily balance ionizations provided that 
$C$ is large. Thus, the larger is $C$, the more gentle is the slope of the 
curve. At $z\approx8.5$ the growth of $Q_{HII}$ temporarily decreases because 
of both the disappearance of Pop~III stars, producing more ionizing photons 
than Pop~II \citep{schaerer2003}, and the quenching of SF in minihaloes. 
The rise of $Q_{HII}$ at $z\approx 7.5$ is due to $10^8\Msun\leq M<10^9\Msun$ 
dwarf galaxies, which become the dominant stellar (photon) sources (Fig.~1a). 
When ${\dot N}_{\gamma}\gg{\dot N}_{rec}$ ionizations completely overcome 
recombinations, and the slope of $Q_{HII}$ becomes independent of $C$. 
The higher is $C$, the later this condition is satisfied, and thus the 
lower is $z_{rei}$. 

The reionization histories obtained by assuming $C=3-5$ (Fig.~2a), provide 
a satisfactory match to the $Q_{HII}$ evolution implied by our choice of 
$M_{sf}(z)$ (blue shaded area). So these models are self-consistent. An 
equally good agreement is achieved by adopting an escape fraction that 
increases with redshift\footnote{Linearly interpolating among the values 
of \cite{mitra2013}: $f_{esc}=0.179$ at $z\geq8$, and $f_{esc}=0.068$ 
at $z\leq6$}, and $C=4$. In this case the MW environment is $\approx65\%$ 
reionized at $z\approx8$, when $M<10^9\Msun$ dwarf galaxies dominate the 
SF (Fig.~1a). Thus, $f_{esc}=0.1$ yields the most conservative results 
for dwarf galaxies.
%%%%%%%%%%%%%%%%%%%%%%%%%%%%%%%%%%%%%%%%%%%%%%%%%%%%%%%%%%%%%%%%%%%%%%%%%%%%%%%%%%%%
\begin{figure}
   \includegraphics[height=0.77\columnwidth]{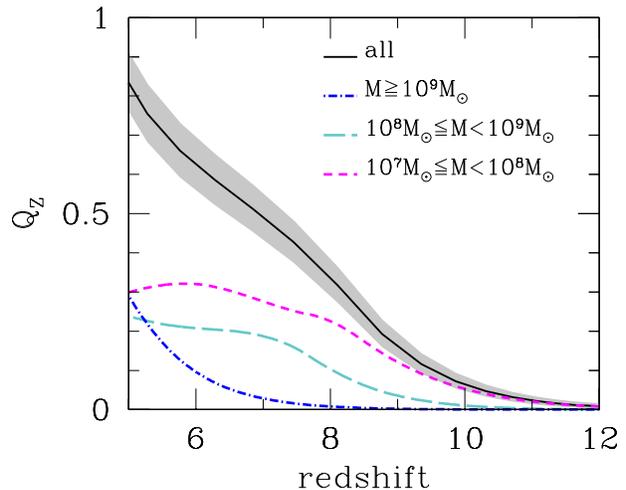}
    \caption{Evolution of the metal filling factor in the MW volume, 
    $Q_Z$, averaged over 50 merger histories (solid line as in Fig.~1a).
    The contribution by galaxies with different masses is shown.}
    \label{Fig4}
\end{figure}
%%%%%%%%%%%%%%%%%%%%%%%%%%%%%%%%%%%%%%%%%%%%%%%%%%%%%%%%%%%%%%%%%%%%%%%%%%%%%%%
Within this assumption we can inspect Fig.~\ref{Fig3} where we show, as a 
function of $z_{rei}$, the contribution of galaxies with different masses 
to the time-integrated comoving photon density, $n_{\gamma}$. We see that
$M<10^9\Msun$ dwarf galaxies are the {\it primary reionization sources} 
of the MW environment, providing $>80\%$ of the ionizing photons 
independent of $z_{rei}$, and thus of $C$. Their contribution rises up to 
$\approx93\%$ for our reference model ($C=3$, $z_{rei}=6.4\pm0.5$). 
At $z\geq8$ the reionization process is driven by $M<10^8\Msun$ dwarf 
galaxies, which provide $>70\%$ of $n_{\gamma}$. At $z\leq8$ the 
photo-dissociating/ionizing radiation produced by the {\it same} 
dwarf galaxies quenches the SF in these systems, and the reionization 
is completed by $10^8\Msun<{M}<10^9\Msun$ dwarf galaxies, which account 
for $\approx55\%$ of the ionizing photon budget at $z_{rei}=6.4$.
%%%%%%%%%%%%%%%%%%%%%%%%%%%%%%%%%%%%%%%%%%%%%%%%%%%%%%%%%%%%%%%%%%%%%%%%%%%%%%%
\subsection{Metal polluters}
%%%%%%%%%%%%%%%%%%%%%%%%%%%%%%%%%%%%%%%%%%%%%%%%%%%%%%%%%%%%%%%%%%%%%%%%%%%%%%%%
The evolution of the metal filling factor, $Q_Z$, is shown 
in Fig.~\ref{Fig4} along with the contribution of galaxies within 
different mass ranges. We find that metals fill $\approx 65\%$ 
of the MW volume by $z\approx 6$, and $\approx 100\%$ by $z\approx 
4.5$, implying that the metal enrichment of the MW environment 
proceeds slower than reionization. In Fig.~2b the $z$-evolution 
of the two filling factors are plotted against each other, and 
we can see that $Q_Z < 1$ when $Q_{HII}\approx 1$, independent 
of the assumed $C$ value. In Fig.~2b we can also see the imprint 
of Pop~III stars, which accelerate the early evolution of $Q_{HII}$ 
but not $Q_Z$. This is because in our model Pop~III stars are 
assumed to form with a standard IMF, and so they have have harder 
spectra then Pop~II but same SN explosion energy. Even with this 
conservative hypothesis we find that at $z=6$ {\it dwarf galaxies} 
can account for $\geq 90\%$ of the metal-enriched volume, and for 
$\geq 65\%$ at $z\approx 5$, when $Q_Z\approx 0.85$. 
At $z=5$ we find that the gas in the MW environment is enriched up 
to an average metallicity $Z_{gm}=M^{Z}_{gm}/M_{gm}\approx 2\times 
10^{-2}\Zsun$ (see also Fig.~4 of SSF07), and that $>90\%$ of these 
metals have been ejected from $M<10^9\Msun$ dwarf galaxies. At lower 
$z$, radiative feedback processes (and merging) strongly reduce the 
SF in dwarf galaxies and thus their instantaneous metal-ejection rate. 
At $z\approx2$ however, $\approx50\%$ of $M^Z_{gm}(z=2)$ still 
originates from these high-$z$ dwarfs, which therefore dominate 
the integrated contribution to the metal-enrichment of the MW 
environment. 
%%%%%%%%%%%%%%%%%%%%%%%%%%%%%%%%%%%%%%%%%%%%%%%%%%%%%%%%%%%%%%%%%%%%%%%%%%%%%%%
\section{Discussion}
%%%%%%%%%%%%%%%%%%%%%%%%%%%%%%%%%%%%%%%%%%%%%%%%%%%%%%%%%%%%%%%%%%%%%%%%%%%%%%%
We used a data-calibrated cosmological model for the formation 
of the Milky Way to show that the high-$z$ counterparts of nearby 
dwarf galaxies could easily be the primary reionization sources and 
the early metal-polluters of the MW environment. Our model predictions 
at $z>5$ are consistent with the SF histories of MW dwarf galaxies: 
ultra-faint dwarfs stop forming stars before the end of reionization 
\citep[][]{brown2012}, which does not affect the evolution of the more 
massive classical dSphs \citep[e.g.][]{monelli2010}. The average SF 
rates predicted for $M<10^9\Msun$ dwarf galaxies are also within the 
observed range for the oldest stellar populations (see Fig.~1b), 
implying that their photon/metal contribution has not been overestimated.

Independent of $z_{rei}$ ($\geq5$), and thus of the parameters $C$
and $f_{esc}$, we found that dwarf galaxies can produce $>80\%$ 
of the ionizing photons required to fully ionize the MW environment, 
$n^{mw}_H\approx{4.5}n^0_{H}$, in agreement with recent models and 
observations for cosmic reionization. 
As \cite{choudhury2008}, we predict that reionization is initially 
driven by $M<10^8\Msun$ haloes, the progenitors of ultra-faint dwarf
galaxies and the most common objects at $z>8.5$. It is then completed
by more massive dwarf galaxies, which are not affected by the 
radiative feedback processes that halt the SF in the smallest systems. 
The same $M<10^9\Msun$ dwarf galaxies that reionize the MW environment 
can also drive its early metal-enrichment via SN-driven outflows 
\citep[see also][]{madau2001}. The metals ejected by dwarf galaxies can 
provide $>90\%$ of the total mass of heavy elements in the MW environment 
at $z\approx 5$, when $Q_Z\approx 0.85$, and $\approx50\%$ at $z\approx 2$. 
This implies that the high-z analogues of {\it present-day} dwarf galaxies 
could have dominated the metal-enrichment of the gaseous environments 
surrounding galaxies for $\approx10$~Gyr, thus suggesting possible links 
between the chemical abundances observed in their ancient stellar populations, 
and in the gas measured in distant QSO absorption lines.
%possible links between the chemical abundance of the gas measured in 
%distant QSO absorption lines, and the heavy elements observed in the 
%ancient stellar populations of present-day dwarf galaxies. 
 
%
%and the heavy elements observed in the ancient stellar populations
%of present-day dwarf galaxies. 
%These findings 
%are qualitatively in good agreement with cosmological 
%hydrodynamic simulations of MW-like galaxies \citep{shen2012}, and
%suggest that there could be a link between the chemical abundance 
%of the gas measured in distant QSO absorption lines, and the heavy 
%elements observed in the ancient stellar populations of present-day 
%dwarf galaxies.

For reasonable values of $C\leq5$ \cite[][]{pawlik2009} the 
reionization histories we found are compatible with our assumed 
evolution of the minimum mass for star-formation, $M_{sf}(z)$, 
which is determined by radiative feedback. Our fiducial model 
($C=3$, $f_{esc}=0.1$), which predicts $Q_{HII}=0.5$ at $z=8\pm1$ 
and $Q_{HII}=0.99$ at $z_{rei}=6.4\pm0.5$, provides an electron 
scattering optical depth that is consistent with WMAP/Planck data,
$\tau_{el}=0.085\pm0.009$.
Assuming that all photons are absorbed by the MW environment 
(mean free path $\lambda_{\rm mfp}\leq1$Mpc), we find that at 
$z\approx6$ the intensity of UV radiation is $J({{\nu}_0})\approx3\times10^{-24}(\lambda_{\rm mfp}/{\mpc})\,$erg$\,$s$^{-1}$Hz$^{-1}$cm$^{-2}$sr$^{-1}$.
Our findings for $z_{rei}$ are consistent with \cite{busha2010}, 
\citep[see also][]{alvarez2009} who constrained $z_{rei}=8^{+3}_{-2}$ 
to match the radial distribution and luminosity function of the MW 
satellites. Thus, although our model cannot make predictions for
how many luminous galaxies survive merging, our inferred history 
of reionization is consistent with these additional constraints.

Finally, we found that Pop~III stars are hosted by low-mass 
dwarf galaxies, and that they no longer form in the MW environment
when $z\lsim 8$. If the Pop~III IMF is biased towards massive stars, 
then the UV emissivity will naturally increase at $z>8$, as required 
by recent models for cosmic reionization. The massive nature of 
primordial stars has been overlooked recently because of the lack 
of a clear signature in the chemical abundances of {\it very} 
metal-poor stars \citep[e.g.,][]{cayrel2004,caffau2011}. These
new observational results, however, remain consistent with the 
predictions of our model for Galactic halo stars, which are 
independent of the Pop~III IMF (SSF07). Thus, we cannot exclude 
that Pop~III stars were massive, as also pointed out by recent 
simulations \citep[$m_{PopIII}\approx40\Msun$][]{hosokawa2011}. 
In conclusion, low-mass dwarf galaxies hosting massive Pop~III 
stars could represent the hidden source of ionizing radiation that is 
missing at high-$z$. Hence their impact on MW environment could be 
even more significant than what found here by using conservative 
assumptions.
%%%%%%%%%%%%%%%%%%%%%%%%%%%%%%%%%%%%%%%%%%%%%%%%%%%%%%%%%%%%%%%%%%%%%%%%%%%%%%%
\section*{Acknowledgements}
We thank R.~Schneider, A.~Skuladottir and the anonymous referee for useful 
comments, and the PRIN-MIUR~2010/11 (prot.~2010LY5N2T) for financial support. 
S.Salvadori was supported by the Netherlands Organization for Scientific 
Research VENI, grant 639.041.233.
%%%%%%%%%%%%%%%%%%%%%%%%%%%%%%%%%%%%%%%%%%%%%%%%%%%%%%%%%%%%%%%%%%%%%%%%%%%%%%%%

%\bibliographystyle{mn2e}
\label{lastpage}
%\bibliography{salvadori}
%\label{lastpage} 
\end{document}